\documentclass[twocolumn,traditabstract,longauth,letter]{aa}
\usepackage{amssymb}
\usepackage{mathptmx}
\usepackage{natbib}
\bibpunct{(}{)}{;}{a}{}{,}  
\usepackage[]{graphicx}
\usepackage{lscape}
\usepackage{txfonts}
\usepackage{url}
\defcitealias{canameras15}{C15}
\defcitealias{canameras17}{C17}
\begin{document}

\title{Planck's Dusty GEMS. III. A massive lensing galaxy with a bottom-heavy stellar initial mass function at \textit{z}=1.5
  \thanks{Based on ALMA data obtained with program 2015.1.01518S and
    VLT data obtained with programs 291.A-5014 and 295.A-5017.}}
\author{R.~Ca\~nameras\inst{1,2,3,4}, N.~P.~H.~Nesvadba\inst{12,3,4}, R.~Kneissl\inst{5,6},  M.~Limousin\inst{7}, R.~Gavazzi\inst{8}, D.~Scott\inst{9}, H.~Dole\inst{1,2,3,4}, B.~Frye\inst{10}, S.~Koenig\inst{11}, E.~Le~Floc'h\inst{12}, I.~Oteo\inst{13,14}} 
\institute{
Institut d'Astrophysique Spatiale, Bat.~121, 91405 Orsay, Cedex, France
\and 
Universit\'e Paris-Sud, France
\and
Universit\'e Paris-Saclay, France
\and
CNRS, France
\and
European Southern Observatory, ESO Vitacura, Alonso de Cordova 3107, Vitacura,
Casilla 19001, Santiago, Chile
\and
Atacama Large Millimeter/submillimeter Array, ALMA Santiago Central Offices,
Alonso de Cordova 3107, Vitacura, Casilla 763-0355, Santiago, Chile
\and
Aix Marseille Universit\'e, CNRS, LAM, Laboratoire d'Astrophysique de Marseille, Marseille, France
\and
Institut d'Astrophysique de Paris, 75014, Paris, UPMC Univ. Paris 6, UMR7095
\and
Department of Physics and Astronomy, University of British Columbia,
6224 Agricultural Road, Vancouver, British Columbia, 6658, Canada
\and
Steward Observatory, University of Arizona, Tucson, AZ 85721, USA
\and 
Chalmers University of Technology, Onsala Space Observatory, Onsala, Sweden 
\and
CEA-Saclay, F-91191 Gif-sur-Yvette, France
\and
Institute for Astronomy, University of Edinburgh, Royal Observatory, Blackford Hill, Edinburgh, EH9 3HJ, UK 
\and
European Southern Observatory, Karl-Schwarzschild-Strasse 2, 85748 Garching, Germany
}
\titlerunning{Ruby's lens. A massive, lensing galaxy at $z=1.5$ with a bottom-heavy IMF}
\authorrunning{Ca\~nameras et al.}  \date{Received / Accepted }

\abstract{We study the properties of the foreground galaxy of the
  Ruby, the brightest gravitationally lensed high-redshift galaxy on
  the sub-millimeter sky as probed by the {\it Planck\/} satellite, and
  part of our sample of Planck's Dusty GEMS. The Ruby
  consists of an Einstein ring of 1.4\arcsec\ diameter at
  $z=3.005$ observed with ALMA at 0.1\arcsec\ resolution, centered on
  a faint, red, massive lensing galaxy seen with HST/WFC3, which
  itself has an exceptionally high redshift, $z=1.525 \pm 0.001$, as confirmed
  with VLT/X-Shooter spectroscopy. Here we focus on the properties of
  the lens and the lensing model obtained with {\tt
    LENSTOOL}. The rest-frame optical morphology of this system is
  strongly dominated by the lens, while the Ruby itself is highly
  obscured, and contributes less than 10\,\% to the photometry out to
  the $K$ band. The foreground galaxy has a lensing mass of
  $(3.70 \pm 0.35) \times 10^{11}\,{\rm M}_{\odot}$.  Magnification
  factors are between 7 and 38 for individual clumps forming two
  image families along the Einstein ring. We present a
  decomposition of the foreground and background sources in the
  WFC3 images, and stellar population synthesis modeling with a
  range of star-formation histories for Chabrier and Salpeter
  initial mass functions (IMFs).  Only the stellar mass range
  obtained with the latter agrees well with the lensing mass.
  This is consistent with the bottom-heavy IMFs of massive
  high-redshift galaxies expected from detailed studies of the
  stellar masses and mass profiles of their low-redshift
  descendants, and from models of turbulent gas fragmentation. This
  may be the first direct constraint on the IMF in a lens at
  $z=1.5$, which is not a cluster central galaxy.}

\maketitle
\section{Introduction}
\label{sec:introduction}

Dark-matter halo mass is a major driver of galaxy evolution
\citep[e.g.,][]{peng10}, but such halo mass is very challenging to 
measure at high redshift. Only in a few exceptional cases is the 
stellar continuum bright enough for absorption-line measurements
\citep[e.g.,][]{vandersande13}. Dynamical masses measured from gas
motions are affected by inclination and beam-smearing effects and,
potentially, the kinetic energy injected by feedback or accretion and
merger events. Moreover, the degeneracy between age, extinction, and
metallicity limits the accuracy of stellar masses estimated from
broadband colors, which is further reduced by the largely unknown
shape of the initial mass function (IMF). Fossil evidence in nearby
massive early-type galaxies suggests bottom-heavy IMFs
\citep[e.g.,][]{conroy13}, whereas galaxy evolution models favor
top-heavy IMFs \citep[e.g.,][]{lacey16}. The lack of direct
constraints on the IMF not only limits the accuracy of stellar mass
estimates in high-redshift galaxies \citep[e.g.,][]{conroy09}, but
also hides from us important information on the regulation mechanisms of
star formation \citep[e.g.,][]{chabrier14}.

The size of the Einstein ring around a massive galaxy only depends on
the projected mass along the line of sight. Gravitational lensing
therefore offers an opportunity to measure directly galaxy masses that are
unaffected by the detailed baryonic properties of the lens. 
The number of known strong lensing galaxies during the major phase of
galaxy evolution at $z\ge1$ is however very small. We are only
aware of two such galaxies at $z\ge1.5$: a massive galaxy at $z=1.62$
found by \citet{wong14}, which is a brightest cluster galaxy
and might therefore have an atypical formation history and mass
profile; and the galaxy found by \citet{vanderwel13} at $z=1.53$, which
unfortunately has no bright line emission and only a photometric
redshift.

Here we characterize a new lensing galaxy at $z=1.525$, which is
magnifying the brightest high-redshift galaxy in the Planck
Catalogue of Compact Sources \citep{planck14xxviii,planck15xxvii}, a
maximally starbursting galaxy at $z=3.0$ (the ``Ruby''; Ca\~nameras et
al.~2017, submitted; C17 hereafter).  The Ruby is part of our follow up 
of the ``Planck's Dusty GEMS'' (gravitationally enhanced submillimeter
sources) sample, which includes the brightest gravitationally lensed 
high-redshift galaxies discovered with {\it Planck\/} on the roughly 
50\,\% of sky not dominated by Galactic foregrounds
\citep[][]{planck15xxvii,canameras15,nesvadba16}. We focus on the
optical and near-infrared properties, including the spectroscopic
redshift of the foreground source and the lensing model, and discuss
the stellar mass and mass-to-light ratio, which favor a bottom-heavy
stellar IMF. The detailed properties of the
background source are discussed in C17.

We adopt the flat $\Lambda$CDM cosmology from \citet{planck14xvi} with 
$H_0=68\,{\rm km}\,{\rm s}^{-1}\,{\rm Mpc}^{-1}$, $\Omega_{\rm m} = 0.31$, and
$\Omega_\Lambda = 1 - \Omega_{\rm m}$. At $z=3.005$ this implies a 
luminosity distance, $D_{\rm L}=26.0\,$Gpc, and a projected physical scale of 
7.9\,kpc\,arcsec$^{-1}$. At $z=1.525$, $D_{\rm L}=11.3\,$Gpc, and the projected 
scale is 8.6\,kpc\,arcsec$^{-1}$. All magnitudes are in the AB system.\vspace{-4mm}

\begin{figure*}
\centering
\includegraphics[height=0.135\textwidth]{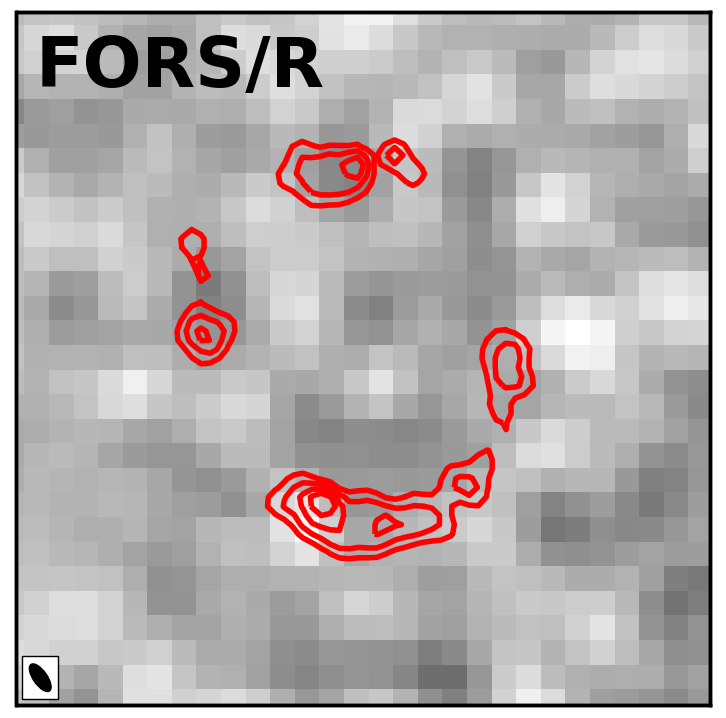}
\includegraphics[height=0.135\textwidth]{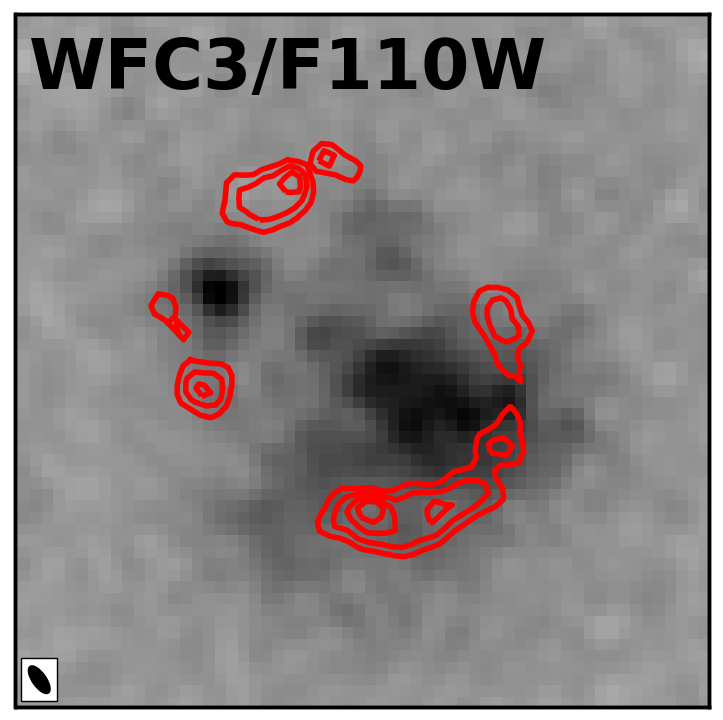}
\includegraphics[height=0.135\textwidth]{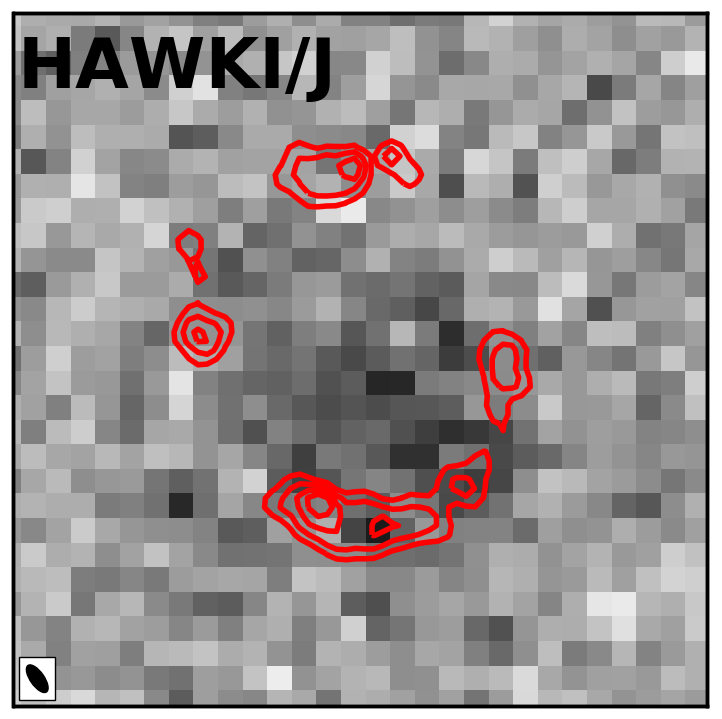}
\includegraphics[height=0.135\textwidth]{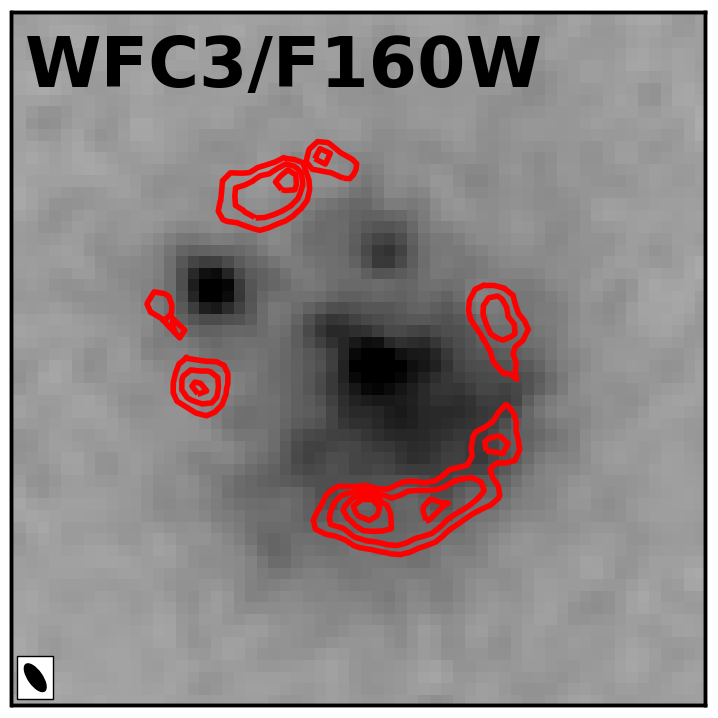}
\includegraphics[height=0.135\textwidth]{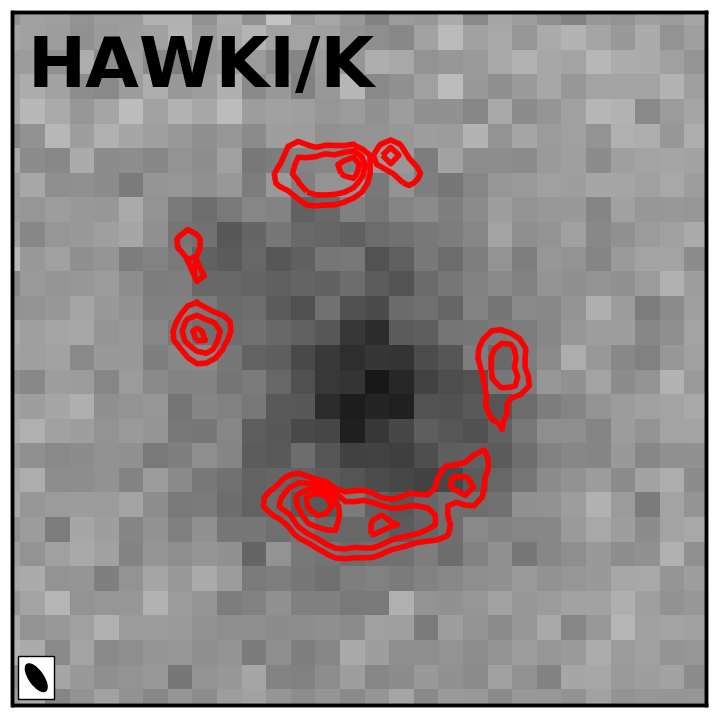}
\includegraphics[height=0.135\textwidth]{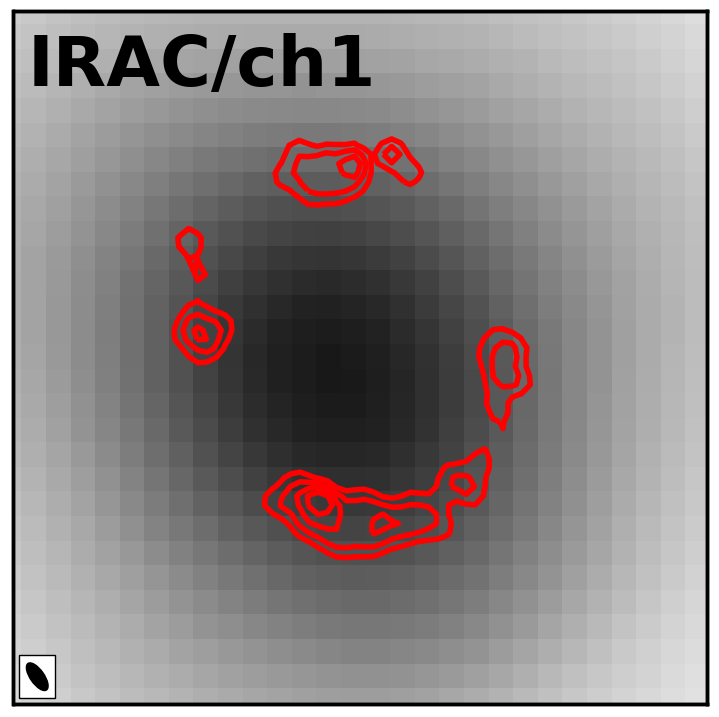}
\includegraphics[height=0.135\textwidth]{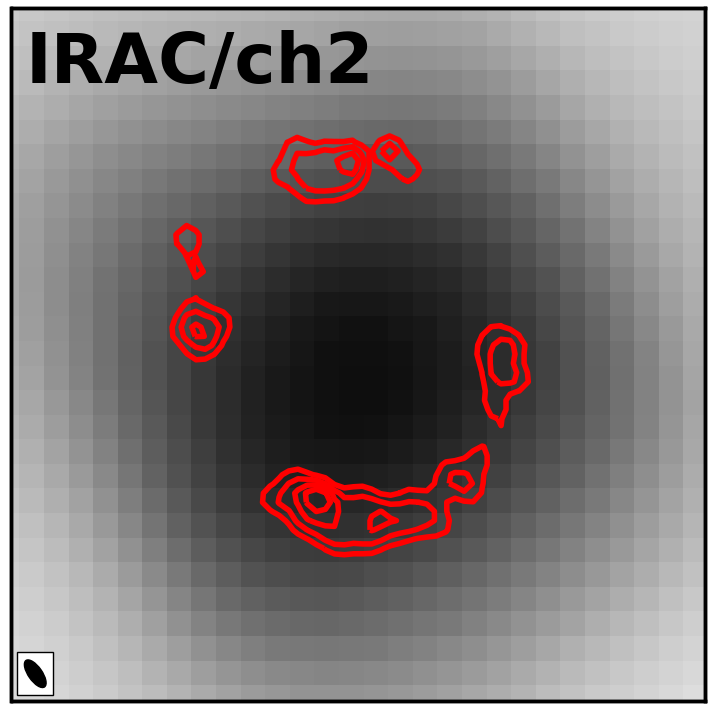}
\caption{{\it Left to right:} Three arcsecond wide postage
  stamps of the Ruby in the $R$, F110W, $J$, F160W, $K$, 3.6-$\mu$m, and
  4.5-$\mu$m bands. Contours show the CO morphology from C17 taken
  with $0.14\arcsec\times0.06\arcsec$ beam size (lower left corner 
  of each panel); they start at 3$\,\sigma$, and increase in steps 
  of 3$\,\sigma$. \vspace{-3mm}}
\label{fig:stamps}
\end{figure*}

\section{Morphology in the near-infrared and millimeter}
\label{sec:observations}

We obtained optical and near-IR imaging of the Ruby on 8 May
2013, using FORS2 through the $V$-, $R$-, and $I$-band filters at
1\arcsec\ seeing with exposure times of 23, 9, and 13\,min and limiting
depths of 27.4, 26.9, and 26.2\,mag, respectively. With HAWK-I we
observed the Ruby on 12 May 2013 with 0.4\arcsec\ seeing, reaching
limiting magnitudes of 24.2, and 23.6\,mag in the $J$ and $K$ band,
respectively, in 26 and 30\,min of total observing time.

We used the ESO pipeline to align the images from the two detectors of
FORS2 and to reduce the images in the standard way, by subtracting
bias or dark frames and dividing by the flatfield.  We used {\sc SWARP}
and {\sc SCAMP} \citep[][]{bertin10a, bertin10b} to align individual
frames relative to each other and within the World Coordinate System,
as probed by the 7th data release of the SDSS, and to
construct the final images. Positional mismatches between bands
suggest an rms uncertainty of at most 0.1\arcsec\ in the final images.
Optical images were flux calibrated using the zero points on the ESO
website and the NIR bands were calibrated relative to 2MASS. We also
ensured that the relative photometry is robust between bands using the
blackbody spectral energy distribution of stars within all bands.

We used high-resolution observations of CO $J\,{=}\,4\,{\rightarrow}\,3$
and dust continuum obtained with ALMA in band~3 (program 2015.1.01518S, 
PI Nesvadba) and with the Wide-Field Camera~3 on the {\it Hubble\/} Space 
Telescope in the F110W and F160W bands (program~14223, PI Frye). These data 
will be discussed in detail in forthcoming papers (C17; Frye et al.~2017, 
in prep.).

The morphology of the Ruby in the optical, near-infrared, and
millimeter is shown in Fig.~\ref{fig:stamps}. The optical and
near-infrared bands are dominated by a faint, red source in the center
of a nearly complete Einstein ring with 1.4\arcsec\ diameter, which can
only be seen at long wavelengths. With ALMA in CO($4{\rightarrow}3$) we detect 
six clumps along this ring (Fig.~\ref{fig:lensmodel}), which we associate
with two systems of multiple images, as described below. \vspace{-4mm}

\section{Spectroscopic redshift with VLT/X-Shooter}
\label{sec:spectroscopy}

We obtained VLT/X-Shooter spectroscopy from 390 to 2500\,nm through DDT
program 295.A-5017 (PI: N. Nesvadba) with a total observing time of
4\,hrs under good and stable conditions and seeing $<1$\arcsec. Slit
widths were 1.2\arcsec\ in the two optical and 1.3\arcsec\ in the NIR
arm. For completeness, earlier X-Shooter data taken in June~2015 and
contaminated by a bright foreground source were discarded. Data were
reduced with the ESO X-Shooter pipeline in the standard way
\citep[][]{modigliani10}.

\begin{figure}
\includegraphics[width=0.45\textwidth]{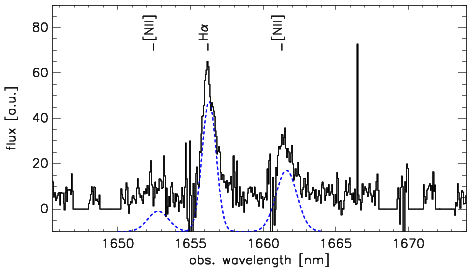}
\caption{VLT/X-Shooter spectrum of H$\alpha$ and
    [\ion{N}{ii}]$\lambda\lambda$6548,6583 in the lens at $z=1.525 \pm 0.001$. Blue
    lines show Gaussian line fits. Bright night-sky line residuals are
    clipped.\vspace{-5mm}}
\label{fig:spectrum}
\end{figure}

We find H$\alpha$ at $(1.6566\,{\pm}\,0.0005)\,\mu$m and the
[\ion{N}{ii}]$\lambda\lambda$6548, 6583 doublet at
$(1.65281\,{\pm}\,0.0005)\,\mu$m and $(1.6617\,{\pm}\,0.0005)\,\mu$m,
respectively, at a common redshift $z=1.525\,{\pm}\,0.001$
(Fig.~\ref{fig:spectrum}). Full width at half maximum (FWHM) line
 widths are $(206\,{\pm}\,20)\,{\rm km}\,{\rm s}^{-1}$ and
$(306\,{\pm}\,41)\,{\rm km}\,{\rm s}^{-1}$ for H$\alpha$, and
[\ion{N}{ii}], respectively. No other line is seen between 0.3 and
2.4\,$\mu$m. Non-detections of [\ion{O}{iii}]$\lambda\lambda$4959, 5007,
H$\beta$, and [\ion{O}{ii}]$\lambda$3728 are consistent with high metallicity
\citep[e.g.,][]{nesvadba07} or extinction; however, the high
[\ion{N}{ii}]$\lambda$6583/H$\alpha$ ratio of $0.69\,{\pm}\,0.1$ suggests the
presence of a faint AGN \citep[][]{kewley13}, which precludes
using [\ion{N}{ii}]/H$\alpha$ as a metallicity indicator. \vspace{-5mm}

\section{Lens modeling and source reconstruction}
\label{sec:lensmodel}

We use {\tt LENSTOOL} \citep[][]{jullo07} to model the strong
gravitational lensing toward the Ruby. {\tt LENSTOOL} is a publicly
available Bayesian lens modeling code, which approximates the
foreground mass distribution through a range of models. We use the
truncated pseudo-isothermal elliptical mass density profiles
\citep[PIEMD;][]{eliasdottir07}.  Briefly summarized, {\tt LENSTOOL}
uses the number of arclets in the image plane, their association with
image systems of the same regions in the source plane, and their
positions relative to the caustic line as input parameters to
constrain the elliptical lens potential and magnification map
\citep[][]{kneib93}. The modeled parameter space is sampled with a
Markov chain Monte Carlo (MCMC) approach, where the posterior
likelihood distribution quantifies the parameter uncertainties.

The near-circular shape of the Einstein ring shows that the faint
source in the center of the ring at $z=1.525$ is the main
deflector. A bright, S0 galaxy at $z=0.13$ is at around
30\arcsec\ distance. The {\tt LENSTOOL} models with and without
this galaxy show that it has no significant impact on the shear and
magnification.

We measure the peak position of individual lensed arclets in the ALMA
CO($4{\rightarrow}3$) flux map at $0.14\arcsec\times0.06\arcsec$ beam
size. The orientation of the velocity gradients indicate the image
parity in the lens plane \citep[see also][]{riechers08,vlahakis15},
but since the velocities partially overlap, identifying image families
unambiguously remains difficult.  We therefore tested 10 plausible
associations of individual arclets into image families and adopted
the single model that converges and places all images at positions
matching those observed. This results in two image systems with two
and four images, respectively, and is a configuration that is also
commonly found in other strong lensing systems
\citep[e.g.,][]{gavazzi08,tu09,limousin09,bayliss11}.  We fitted the
centroid position in right ascension, $\Delta{\rm RA}$, and
declination, $\Delta{\rm Dec}$, the velocity dispersion of the PIEMD,
$\sigma_{\rm PIEMD}$, as well as its ellipticity, $\epsilon$, and
position angle, PA \citep[see also][]{limousin13,bonamigo15}.  The
core and cutoff radii are poorly constrained with an Einstein ring of
about 6.2\,kpc at $z=1.525$. We therefore choose to fix
$r_{\rm core}=0.15\,$kpc \citep[e.g.,][]{limousin07b,richard14}, and
$r_{\rm cut}=100\,$kpc \citep[e.g.,][]{brainerd96}, which are typical 
values for isolated galaxies. Varying these values within our 
observational constraints changes the mass estimate by $<10$\,\%.

\begin{figure}
\centering
\includegraphics[height=100pt]{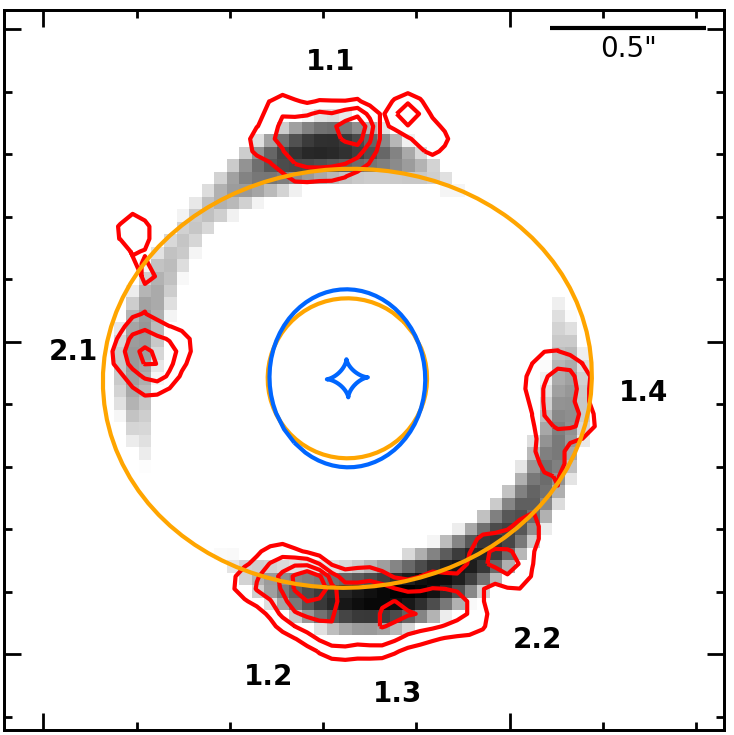}
\includegraphics[height=100pt]{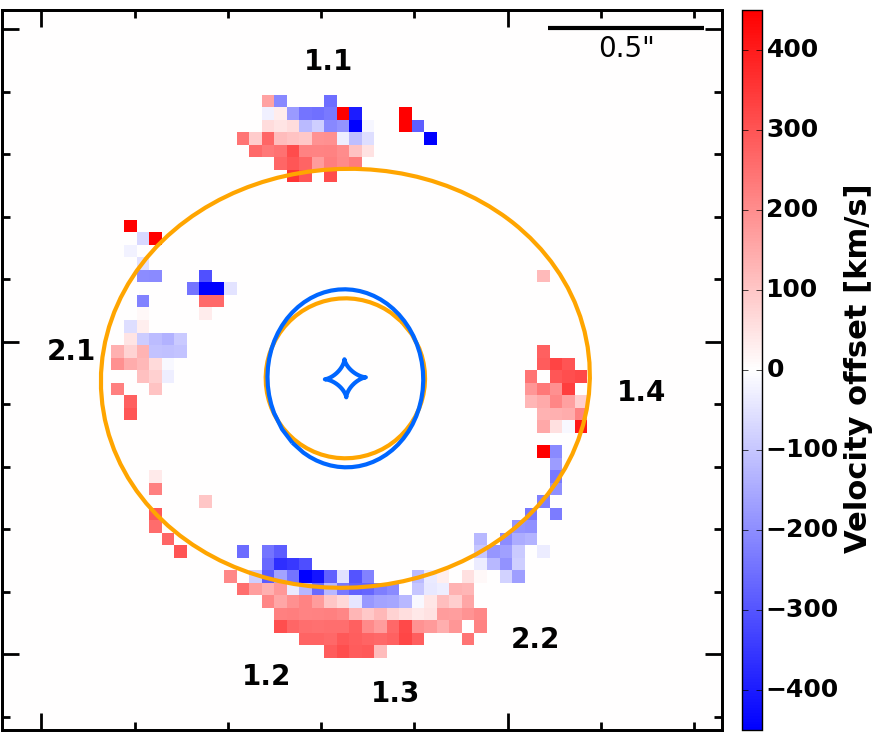}
\caption{{\it Left:} Reconstructed (grayscale) and observed
  (contours) CO($4{\rightarrow}3$) surface brightness observed with ALMA. {\it
    Right:} CO($4{\rightarrow}3$) velocity map of the Ruby. We label the two sets
  of multiple images; we show the tangential and
  radial critical curves at $z=1.525$ and caustic lines at $z=3.005$
  as orange and blue lines, respectively.\vspace{-4mm}}
\label{fig:lensmodel}
\end{figure}

With this approach, we find that the data are best fitted with a
potential of ellipticity $\epsilon=0.113$, position angle
${\rm PA}=+2.7^\circ$ (measured from north to east), and velocity dispersion
$\sigma_{\rm PIEMD}=260.6\,{\rm km}\,{\rm s}^{-1}$. The reconstructed center of
mass is offset by $\Delta{\rm RA}=-0.054\arcsec$ and
$\Delta{\rm Dec}=+0.100\arcsec$, from the reference position at
$\alpha=163.47138^\circ$ and $\delta=5.938551^\circ$, respectively.
Fig.~\ref{fig:lensmodel} illustrates that the source falls near the
tangential critical line and that the data are well fitted by our
model. The rms of the offsets between observed and reconstructed
images is 0.07\arcsec, which is about half the ALMA beam. Varying
the image positions within this beam does not significantly alter the
results.  The best-fitting PIEMD is offset by
$\la0.1\arcsec$ from the galaxy center in the WFC3/F160W
image. At low redshift, we would not expect such offsets
\citep{koopmans06}; but at $z=1.5$, this offset could be from
variations in extinction or star formation history, if not astrometric
uncertainties.

We compute a mean magnification map and uncertainties for each pixel
using 4000 MCMC iterations. Table~\ref{tab:g244p8_ampli} lists the
magnification factors at the peak of each image and the range in
magnification in each spatially resolved (C17) image. In small regions
around caustics factors are up to 60--100. Magnification factors are
consistent with flux ratios. 

The source-plane morphology of each image is reconstructed by
ray tracing the image pixel by pixel through the modeled gravitational
potential \citep[``cleanlens'';][]{sharon12}. We use again 4000 MCMC
iterations to determine the uncertainties on the source-plane
positions. We obtain matching positions for each component of the two
image systems, near the diamond-shaped caustic, offset by 470\,pc from
each other in the source plane. Their intrinsic properties are further
discussed in C17. \vspace{-5mm}

\section{A massive lensing galaxy at $z=1.5$}
\label{sec:decomposition}

Gravitational lensing provides the most accurate estimates of the
integrated mass within the Einstein radius. Following
\citet{limousin05}, we derive the mass of the deflector of the Ruby
from the mean radial profile of the PIEMD and the parameters
listed in Sect.~\ref{sec:lensmodel}, finding an enclosed mass of
$M_{\rm aper}(\theta_{\rm E}) = (3.70 \pm 0.35)
\times 10^{11}\,{\rm M}_{\odot}$ that is
accurate to about 10\,\%, within the Einstein ring, $\theta_{\rm E} =
0.72\arcsec \pm 0.06\arcsec$, corresponding to $(6.2 \pm 0.5)\,$kpc at
$z=1.525$.

\begin{figure}
\centering
\includegraphics[width=0.45\textwidth]{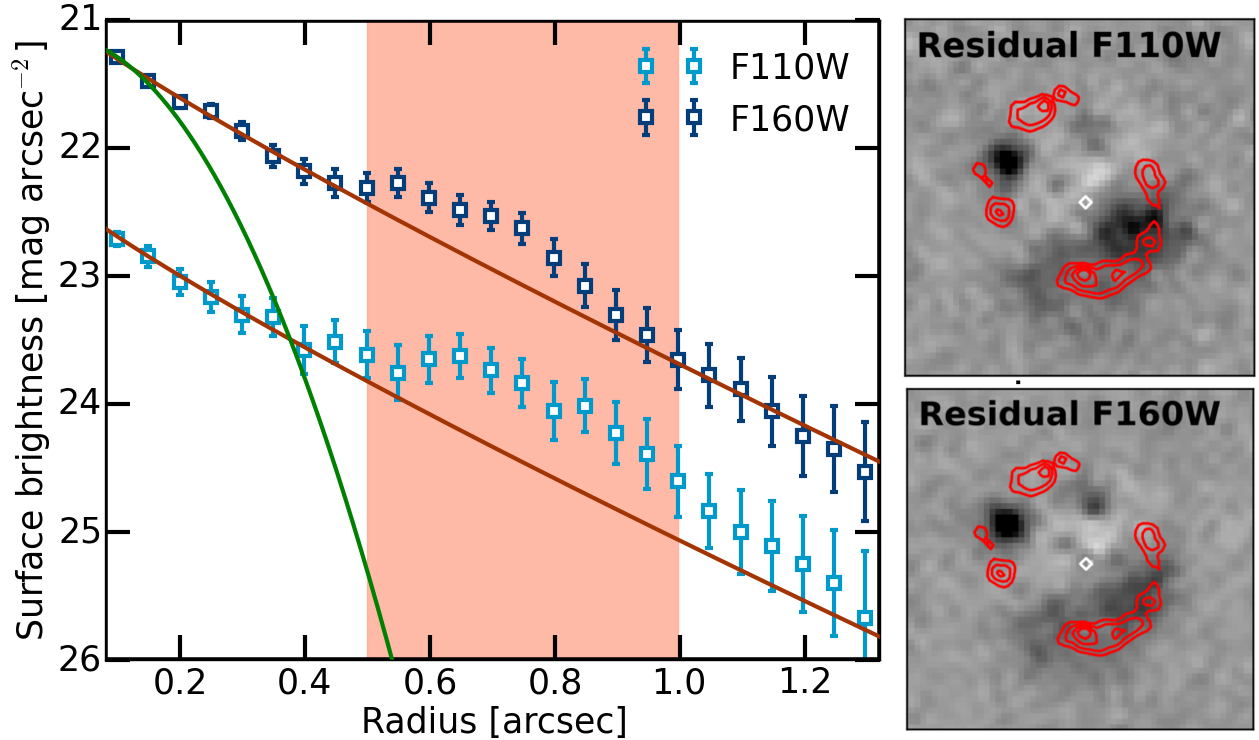}
\caption{{\it Left:} Azimuthally-averaged surface brightness profiles
  in the WFC3/F160W and 110W bands. The PSF is shown in green. The red
  band indicates radii near the critical line, where the Ruby might
  contribute and where an excess is seen compared to the Sersic
  profile (red line) fitted to the inner 0.5\arcsec\ in the
  F160W band. The fit to the F110W band was simply scaled to the
  central surface brightness. These regions are also seen in the
  residual images ({\it right}) after subtracting the model.\vspace{-5mm}}
\label{fig:deblend}
\end{figure}

We can estimate a stellar mass of the deflector from our optical and
NIR photometry, after constraining the possible contamination from the
Ruby. To do this we calculate an azimuthally averaged
surface-brightness profile from our highest resolution image,
WFC3/F160W (Fig.~\ref{fig:stamps}), after masking a ring with a
thickness of twice the PSF (0.18\arcsec) around the caustic line (red
band in Fig.~\ref{fig:deblend}) to avoid contamination from the
Ruby. Fig.~\ref{fig:deblend} shows that the deflector is well
fit with a near exponential profile that has a Sersic index of
$1.15\pm0.29$ and a half-light radius of $0.75\arcsec\,{\pm}\,
0.23\arcsec$, or $(6.5\pm1.9)\,$kpc at $z=1.525$. We do
not see strong residuals in the inner region, but observe a faint
excess at radii near the caustic line, corresponding to about 15\,\%
of the total flux. The same profile is also a good match to the
WFC3/F110W image (Fig.~\ref{fig:deblend}). Convolving this profile
with the HAWK-I $K$-band point spread function suggests that the
contamination from the Ruby in this band is also very mild, indicating
that the optical and NIR photometry can be used to constrain the
star formation history and stellar population in the deflector.

We use the population synthesis code of \citet{bruzual03}, with solar
metallicity, as is common for massive galaxies, a \citet{chabrier03}
IMF, and the \citet{calzetti94} extinction law to
fit the photometry of the deflector in the $I$, F110W, F160W, and $K$
band with an exponentially declining star formation history with
$\tau=20$--200\,Myr, and ages between 1 and 3\,Gyr and
$A_V=1$--2\,mag. Stellar absorption line spectra in massive
early-type galaxies at low and high redshift with enhanced abundance
ratios of $\alpha$ elements relative to iron suggest bursty
star formation histories for massive galaxies like the one studied here
\citep[][]{thomas05,kriek16}. Younger unobscured starbursts are
ruled out by the absence of a blue continuum and older ages by the
age of the Universe at $z=1.5$. A young, dusty burst is ruled out by
the upper limit on the dust continuum at 3\,mm from our ALMA data (C17)
of 0.14\,mJy rms, implying star formation rates $<120$--140\,$M_{\odot}\,
{\rm yr}^{-1}$. Our best-fitting model that does not overestimate the
emission in the optical is 3\,Gyr old with $A_v=1.5$\,mag. Overall,
models within the above range of star formation histories fitting our
SED with $\chi^2<5$ result in stellar masses of 2--$2.5
\times 10^{11}\,{\rm M}_\odot$. When using a \citet{salpeter55} IMF
instead, we find masses that are greater by a factor of 1.7. These
estimates do not depend strongly on metallicity. For example, when
using models with 2.5 times the solar metallicity, we found stellar
masses that were lower by 12--15\,\%.

\begin{figure}
\centering
\includegraphics[height=0.28\textwidth]{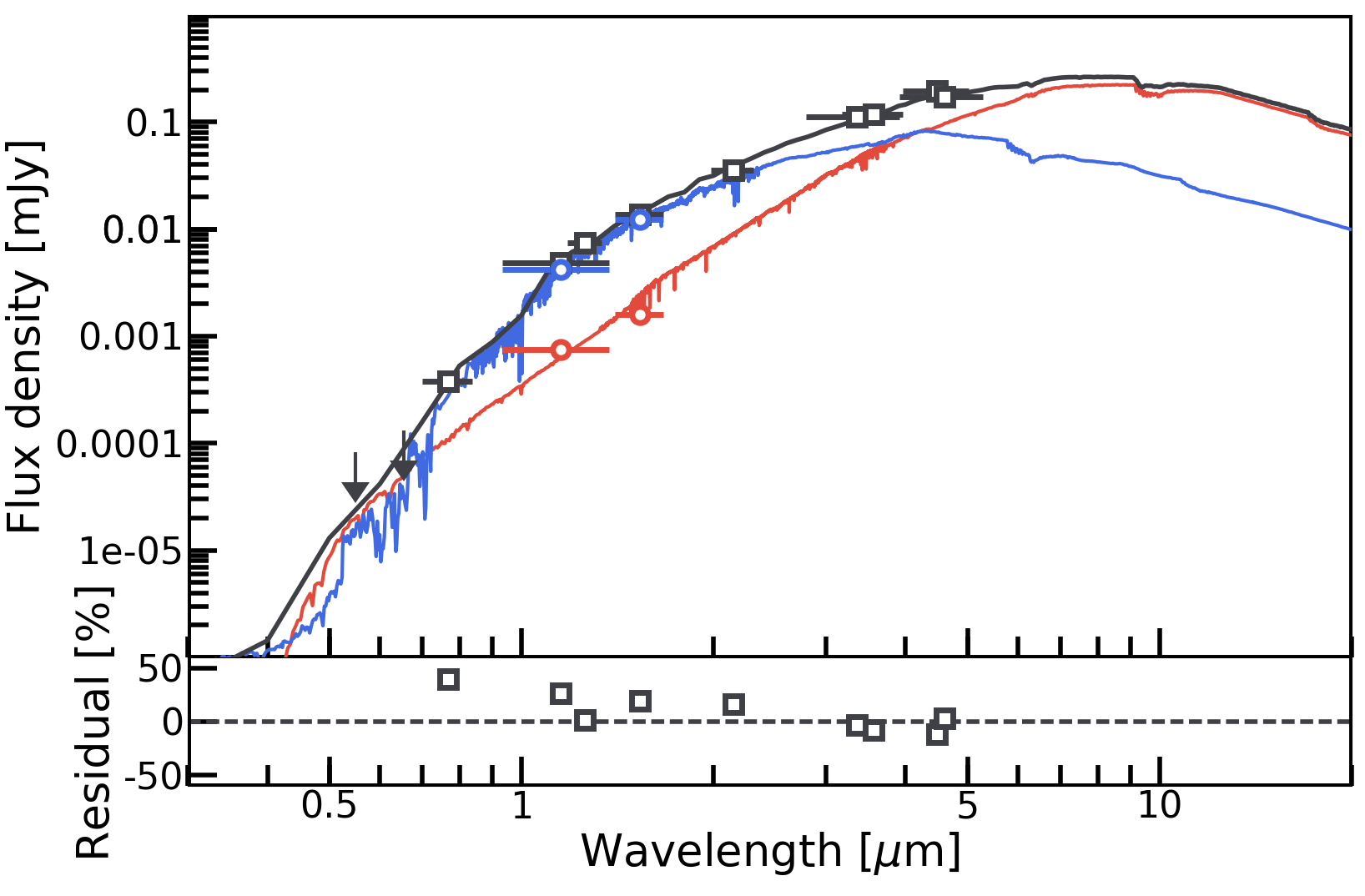}
\caption{Spectral energy distribution of the deflector (blue) and Ruby (red). 
  The combined SED is shown in black. Black squares show the total photometry, 
  while blue and red dots are the resolved photometry of the foreground galaxy 
  and the Ruby in the F110W and F160W images, respectively.\vspace{-2mm}}
\label{fig:stellar_pop}
\end{figure}

In Fig.~\ref{fig:stellar_pop} we show the optical-to-NIR photometry of
the Ruby and our best-fit SED of the lens, matched to the ground-based
photometry from the $V$ to $K$ bands. In the IRAC 3.6-$\mu$m and
4.5-$\mu$m bands the contribution from the Ruby is no longer
negligible (Fig.~\ref{fig:stellar_pop}). The combined photometry for
both sources is well matched, when adopting a young burst with
$\tau=50$\,Myr, 10\,Myr age, and $A_V=4.5$\,mag for the Ruby.

\vspace{-5mm}
\section{Evidence for a bottom-heavy IMF}
The mass profiles of massive galaxies at low redshift suggest that
dark matter contributes at most 10\,\% to the total mass out to 6--7\,kpc
\citep{auger10,conroy13} and probably less in our source at $z=1.5$,
owing to gradual dark-matter assembly \citep[][]{maccio08}. The
non-detection of the dust continuum with ALMA (C17) also rules out
masses of cold molecular and atomic gas above on the order of $10^9\ {\rm
  M}_{\odot}$ (1\,\% of $M_{\rm lens}$), for a simple
scaling with the range of gas-to-dust ratios between 40 and 100, as
measured by \citetalias{canameras15}. This suggests $M_{\rm
  stellar}\simeq M_{\rm lens}$, which is the case for a Salpeter IMF,
whereas a Chabrier IMF underpredicts the lensing mass by about a
factor of 2. Flux near the Einstein ring contributes only 15\,\% to the
total flux in the WFC3 images, and therefore does not change our
result. We also ruled out the IMFs with average mass-to-light ratios 
that are much greater than Salpeter, 0.4--$1.5\,{\rm M}_{\odot}/{\rm L}_{\odot}$ 
in the rest-frame $V$ band, which would produce masses greater than the 
lensing mass.

The difference between the two IMFs is the slope at stellar masses
$<1\,{\rm M}_{\odot}$, where \citet{chabrier03} predict a shallower slope
than the single slope of $\alpha=-2.35$ adopted by \citet{salpeter55},
and hence more bottom-light IMFs. Such low-mass stars are difficult
to observe directly. However, the best estimates currently available,
including those based on direct spectral tracers of low-mass stars,
favor bottom-heavy IMFs consistent with Salpeter or even steeper
slopes \citep[][]{auger10, conroy13, sonnenfeld17, vandokkum16}, which are
also consistent with our results. These IMFs should scale with galaxy
mass, but not with redshift \citep[][]{sonnenfeld17}; this is consistent with
mainly passively evolving stellar populations since redshifts $z=1$--2.

The low-mass end of the IMF in high-redshift galaxies is a sensitive
probe of feedback processes in massive starburst galaxies. For example,
\citet{larson05} finds that increased Jeans masses in dense gas could
make the IMF more top-heavy, whereas \citet{chabrier14} argue that
turbulent gas fragmentation should produce bottom-heavy IMFs similar, 
or even steeper than Salpeter. Our result favors these bottom-heavy IMFs, 
and thus exacerbates the long-standing difficulty of semi-analytic 
galaxy evolution models to match the predicted upper end of the galaxy 
mass function and number of submillimeter galaxies with observations 
without imposing a top-heavy IMF \citep[][]{lacey16}. It will therefore 
be very interesting to see whether our results are representative of the 
overall population of massive high-$z$ galaxies, as more massive, strong 
lensing galaxies become known at these redshifts. \vspace{-5mm}

\section*{Acknowledgments}
We thank the referee for comments that helped improve the paper, the
ESO Director General for granting DDT, and the staff at Paranal, ALMA,
and NASA for carrying out the observations. RC would like to thank
C. Grillo for helpful discussions. ML acknowledges CNRS and CNES for support.
OI acknowledges support from ERC program 321302, {\sc COSMICISM}. 
This paper makes use of ALMA data ADS/JAO.ALMA\#2015.1.01518.S. \vspace{-4mm}

\bibliography{g244p8}

\begin{table}
\centering
\begin{tabular}{lcccccc}
\hline
\hline
ID & 1.1 & 1.2 & 1.3 & 1.4 & 2.1 & 2.2 \\
\hline
$\mu_{\rm clump}$ & $22\pm2$ & $41\pm5$ & $30\pm3$ & $17\pm3$ & $11\pm2$ & $42\pm6$\\
$\mu_{\rm peak}$ & $11\pm1$ & $27\pm3$ & $11\pm1$ & $13\pm2$ & $\phantom{1}7\pm1$ & $22\pm2$ \\
\hline
\end{tabular}
\caption{Gravitational magnification factors; $\mu_{\rm clump}$ lists
  luminosity-weighted average magnifications in each image. $\mu_{\rm
    peak}$ is the magnification at the peak.\vspace{-8mm}}
\label{tab:g244p8_ampli}
\end{table}

\end{document}